\documentclass[12pt]{article}
\renewcommand{\baselinestretch}{1.25}
\usepackage{graphicx}
\begin{document}
%
\begin{flushright}
  OU-HET-501 \ \\
\end{flushright}
\vspace{0mm}
\begin{center}
\large{Chiral-Odd Twist-3 Distribution Function $e(x)$}
\end{center}
\vspace{0mm}
\begin{center}
M.~Wakamatsu and Y. Ohnishi
\end{center}
\vspace{-5mm}\begin{center}
Department of Physics, Faculty of Science, \\
Osaka University, \\
Toyonaka, Osaka 560-0004, JAPAN
\end{center}

\vspace{-2mm}
\begin{center}
\small{{\bf Abstract}}
\end{center}
\vspace{0mm}
\begin{center}
\begin{minipage}{15.5cm}
\renewcommand{\baselinestretch}{1.0}
We clarify the nonperturbative origin of the $\delta$-function
singularity at $x = 0$ in the chiral-odd twist-3 distribution
function $e(x)$ of the nucleon. We also compare a theoretical
prediction for $e(x)$ based on the chiral quark soliton model
with the empirical information extracted
from the CLAS semi-inclusive DIS measurement.
\end{minipage}
\end{center}

\vspace{3mm}
\section{Introduction}

The subject of my talk here is the spin-independent chiral-odd
twist-3 distribution function $e(x)$ of the nucleon.
Why is it interesting? Firstly, its first moment is proportional
to the famous $\pi N$ sigma term. Secondly, within the framework
of perturbative QCD, it was pointed out by Burkardt and Koike that
this distribution function is likely to have a delta-function type
singularity at $x=0$ \cite{BK2002}.
Unfortunately, the physical origin of this
singularity was not fully clarified within the perturbative analysis.

The purpose of my present talk is two-fold. First, I want to show
that the physical origin of this peculiar singularity is inseparably
connected with the {\it nontrivial vacuum structure of QCD}.
Secondly, I will show some theoretical predictions for this interesting
quantity based on the Chiral Quark Soliton Model (CQSM), which can,
for example, be compared with the recent CLAS measurement of
semi-inclusive DIS scatterings \cite{CLAS}.

\section{Origin of delta-function singularity in $e(x)$}

We start with the general definition of $e(x)$ given as
\begin{equation}
 e(x) \ = \ M_N \,\int_{- \infty}^{\infty}
 \,\frac{d z^0}{2 \pi} \,e^{- i x M_N z^0}
 \,E(z^0) ,
\end{equation}
with
\begin{equation}
 E(z^0) \ = \ \langle N \vert \,
 \bar{\psi}(- \frac{z}{2}) \,\psi (\frac{z}{2}) \,\vert
 N \rangle {|}_{z_3 = - z_0, z_\perp = 0} .
\end{equation}
Here, the quantity $E(z_0)$ as a function of $z_0$ measures
light-cone quark-quark correlation of scalar
type in the nucleon. The existence of delta-function singularity
in $e(x)$ indicates that, when $z_0$ becomes large, $E(z_0)$ does
not damp and approaches a certain constant such that
\begin{equation}
 E(z^0) \ = \ E^{reg}(z_0) \ + \ \mbox{constant}, \label{asymp}
\end{equation}
with $E^{reg}(z^0) \rightarrow 0$ as $z^0 \rightarrow \infty$.
In fact, within the framework of the CQSM, we have analytically
confirmed this behavior \cite{WO2003}. As proved there, the
existence of this infinite-range correlation is inseparably
connected with the nontrivial vacuum structure of QCD, or more
concretely, the appearance of nonvanishing quark condensate.

\vspace{6mm}
\begin{figure}[ht]
\begin{center}
 \includegraphics[width=8.0cm]{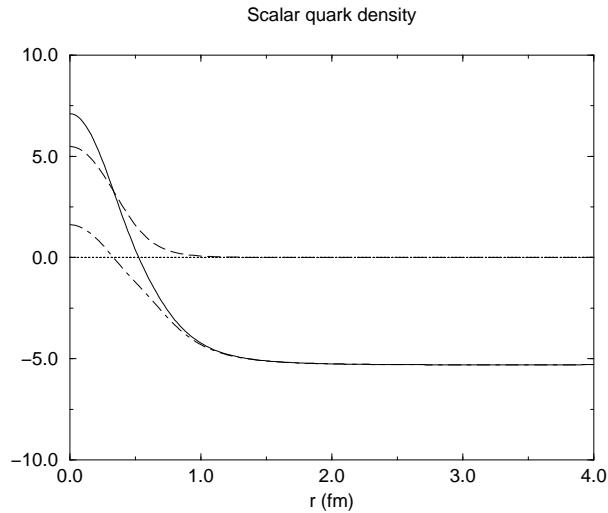}
\end{center}
\vspace*{-0.5cm}
\caption{The scalar quark density predicted by the CQSM.}
\end{figure}
\vspace{4mm}

A natural question is ``Why the vacuum property comes into a localized
hadron observable?". This is due to an extraordinary nature of the
scalar quark density inside the nucleon illustrated in Fig.1. (This
is the prediction of the CQSM.)
The long-dashed curve represents the contribution of 3 valence
quarks in the hedgehog mean field, while the dash-dotted curve stands
for the contribution of deformed Dirac-sea quarks. One clearly sees
that, as the distance from the nucleon center becomes large, the
valence quark contribution damps rapidly, while the Dirac-sea
contribution does not damp, and approaches some {\it negative constant},
which is nothing but the {\it vacuum quark condensate}.
The existence of unending stretch of the region with nonzero scalar
quark density is inseparably connected with the peculiar behavior of
$E(z^0)$ as given by (\ref{asymp}).

\section{Numerical study of $e(x)$ in CQSM}

Now, we turn to the numerical study of $e(x)$ within the CQSM.
In this theory, the isoscalar and isovector combinations of $e(x)$
have very different $N_c$ dependence. The former is an order $N_c$
quantity, while the latter a subleading quantity in $N_c$.
We need very sophisticated numerical method to handle the isoscalar
distribution function $e^{(T=0)}(x)$ containing a delta-function
type singularity. The detail can be found in [4].
After all, we find that
the total $e^{(T=0)}(x)$ is given as a sum of the two terms, i.e.
the the valence quark contribution, which has a peak around
$x = 1/3$, and the Dirac-sea contribution as
\begin{equation}
 e^{(T=0)}(x) \ = \ e^{(T=0)}_{valence}(x) \ + \ 
 e^{(T=0)}_{sea}(x) .
\end{equation}
The latter is further decomposed into two terms :
\begin{equation}
 e^{(T=0)}_{sea}(x) \ = \ e^{(T=0)}_{sing}(x) \ + \ 
 e^{(T=0)}_{regl}(x) .
\end{equation}
Here, the first is a singular term proportional to a delta function,
\begin{equation}
 e^{(T=0)}_{sing}(x) \ = \ C \,\delta(x), \ \ \ \ \ C \ \simeq \ 9.9,
\end{equation}
while the second is a regular term, the magnitude of which turns out
to be relatively small.

Particularly interesting here is the 1st moment sum rule for the
isoscalar $e(x)$, which gives the nucleon scalar charge
$\bar{\sigma}$, i.e. $\int_{-1}^1 \,e^{(T=0)}(x) \,dx \ = \ \bar{\sigma}$.
Numerically, we find that
\begin{equation}
 \bar{\sigma} \ = \  \bar{\sigma}_{valence} 
 \ + \ \bar{\sigma}_{sea}^{regl}
 \ + \ \bar{\sigma}_{sea}^{sing} 
 \ \simeq \  1.7 \ + \ 0.18 \ + \ 9.92 \ \simeq \ 11.8 ,
\end{equation}
which means that the singular term gives
dominant contribution to this sum rule. With the standard
values of the current quark mass $m_q \simeq (4 \sim 7) \,\mbox{MeV}$,
this nucleon scalar charge gives fairly large $\pi N$ sigma term
\begin{equation}
 \Sigma_{\pi N} \ \equiv \ m_q \,\bar{\sigma} \ \simeq \ 
 (47 \sim 83) \,\mbox{MeV} ,
\end{equation}
which favors the recent analyses of low-energy $\pi N$ scattering
amplitudes \cite{PIN2002}.

For the isovector part of $e(x)$, we just comment that no
singularity at $x=0$ is observed. This is reasonable since
there is no isovector quark condensate in the QCD vacuum.
Combining the isoscalar and isovector distributions, we can
predict any of the $u, d, \bar{u}, \bar{d}$ distributions.
Fig.2 shows the comparison of the predicted flavor combination
$e^u (x) + (1/4) \,e^{\bar{d}}(x)$ with the empirical information
extracted from the CLAS data by Efremov et al. under the
assumption of Collins mechanism dominance \cite{EGS2002}.
One sees that the agreement between the theory and experiment
is encouraging, although it would be premature to extract any
decisive conclusion only from this crude comparison on the possible
violation of the pion-nucleon sigma term sum rule due to the existence
of the delta-function singularity at $x=0$.

\vspace{5mm}
\begin{figure}[ht]
\begin{center}
 \includegraphics[width=8.0cm]{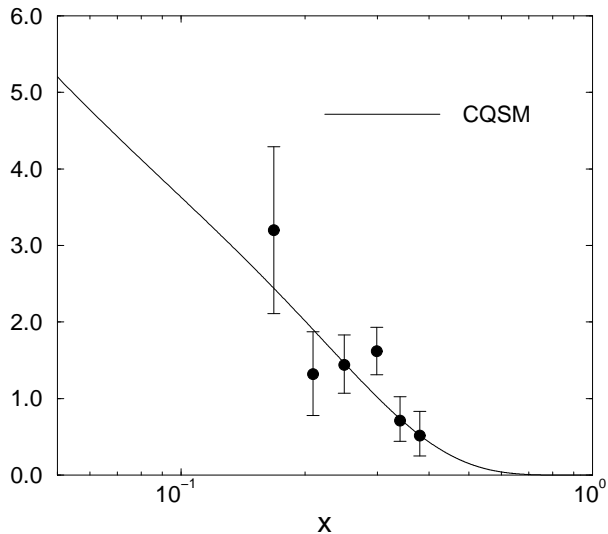}
\end{center}
\vspace*{-0.5cm}
\caption{Comparison with a parameterization of the CLAS data.}
\vspace{10mm}
\end{figure}

\section{Summary and Conclusion}

Summarizing my talk, I have shown that the delta-function singularity
in the chiral-odd twist-3 distribution $e(x)$ is a manifestation of the
nontrivial vacuum structure of QCD in a hadron observable.
Experimentally, the existence of this singularity will be observed as
the violation of $\pi N$ sigma-term sum rule of isoscalar $e(x)$.
To confirm this interesting possibility, we certainly need more
precise experimental information for $e(x)$ especially in the
smaller $x$ region.

\end{document}